\PassOptionsToPackage{square,comma,numbers,sort&compress,super}{natbib}

\documentclass{article}

\usepackage{arxiv}

\usepackage[utf8]{inputenc} % allow utf-8 input
\usepackage[T1]{fontenc}    % use 8-bit T1 fonts
\usepackage{hyperref}       % hyperlinks
\usepackage{url}            % simple URL typesetting
\usepackage{booktabs}       % professional-quality tables
\usepackage{amsfonts}       % blackboard math symbols
\usepackage{nicefrac}       % compact symbols for 1/2, etc.
\usepackage{microtype}      % microtypography
\usepackage{lipsum}
\usepackage{amsmath}
\usepackage{graphicx}
\usepackage{subfigure}
\usepackage{tabularx}
\usepackage{array}
\usepackage{multirow}
\usepackage{caption}
\usepackage[title,toc,titletoc,header]{appendix}

\title{A Probabilistic Approach to Identifying  
Run Scoring Advantage in the Order of Playing Cricket}

\author{
  Manar D. Samad \\
  Department of Computer Science\\
  Tennessee State University\\
 Nashville, TN, USA \\
  \texttt{msamad@tnstate.edu} \\
  \and
   {\bf Sumen Sen} \\
  Department of Mathematics and Statistics\\
 Austin Peay State University\\
 Clarksville, TN, USA  \\
}

\begin{document}

\maketitle

\begin{abstract}
In the game of cricket, the result of coin toss is assumed to be one of the determinants of match outcome. The decision to bat first after winning the toss is often taken to make the best use of superior pitch conditions and set a big target for the opponent. However, the opponent may fail to show their natural batting performance in the second innings due to a number of factors, including deteriorated pitch conditions and excessive pressure of chasing a high target score. The advantage of batting first has been highlighted in the literature and expert opinions, however, the effect of batting and bowling order on match outcome has not been investigated well enough to recommend a solution to any potential bias.  This study proposes a probability theory-based model to study venue-specific scoring and chasing characteristics of teams under different match outcomes. A total of 1117 one-day international matches held in ten popular venues are analyzed to show substantially high scoring advantage and likelihood when the winning team bat in the first innings. Results suggest that the same ‘bat-first’ winning team is very unlikely to score or chase such a high score if they were to bat in the second innings. Therefore, the coin toss decision may favor one team over the other. A Bayesian model is proposed to revise the target score for each venue such that the winning and scoring likelihood is equal regardless of the toss decision. The data and source codes have been shared publicly for future research in creating competitive match outcomes by eliminating the advantage of batting order in run scoring. 
\end{abstract}


\section{Introduction}

The game of cricket has more than two billion fans and followers over the world with 104 cricket playing nations. One-day international (ODI) and T-20 formats are the most popular versions of cricket that are played between teams in home-away series, world-cup tournaments, and domestically at first class tournaments and premium leagues. In the game of cricket, one team (Team A) bat first (first innings) to score 'runs' by competing against the bowling and fielding performance of the opponent team (Team B). In ODI, eleven players (ten wickets) of team A bat in the first innings to score a total ‘run’ facing 50 overs (300 balls, 6 balls per over) of bowling delivery of the opponent team B. Following a half-time break, the team B bat in the second innings to chase the target score playing against  50 overs of bowling delivery of team A. Team A win, tie, or lose the match if team B score less, equal, or more than the first innings target score set by Team A, respectively. The same strategy is followed in the T-20 version of the game, but each team is given only 20 overs (120 balls) to score or chase instead of 50 overs.

Unlike other popular sports such as soccer, hockey, basketball, the game of cricket is unique that involves heavy accounting and diverse statistics to evaluate or predict the performance of a team or individual players. Statistical analyses may play a valuable role in determining an effective game strategy and team selection~\citep{Amin2014}, analyzing outcome in the second innings based on target set in the first innings~\citep{Bose2019}, assessing the performance of players~\citep{Akhtar2015}, ordering of eleven players during the batting session~\citep{Swartz2006}, and predicting outcome of a match~\citep{Akhtar2012, Asif2019}.

Therefore, historical data obtained from thousands of ODI cricket matches along with data-driven statistical methods can be valuable resources for operational research and management, performance evaluation, updating game rules, and forecasting results~\citep{Mondin2012}. One of the most popular statistical models adopted in Cricket is the Duckworth-Lewis (DL) method that determines a revised and fair target score when the game is interrupted, and the match duration is shortened by rain~\citep{Duckworth1998}. In the past, the event of rain not only postponed the match, but also unfairly penalized one of the teams by cutting their allotted overs for batting. The DL method has been used for over two decades to set a reasonable target by statistically considering partial match results and resources (wickets and overs) available prior to the rain. This method has been an active subject of research and modification over the last two decades~\citep{McHale2013, Preston2002}.

Apart from rain interrupted events, there are numerous implicit sources of bias that may unfairly favor one team over another~\citep{Forrest2008}. For example, the home ground advantage is inevitable that is likely to favor the hosting team in a cricket match~\citep{Morley2005}. The innings played at night time in a day-night match has shown considerable difference in outcome when compared to its counterpart innings played in the day time~\citep{McGinn2013}.  One of the debated issues in Cricket is the decision whether a team should bat or bowl first following a toss of a fair coin. This decision has been argued to give an advantage to the toss winning team. Sood and Willis have shown in a recent study that the winning of coin toss has a significant effect on winning the game, especially when the contesting teams have matching performance, and the match is played in certain conditions such as in the day-night format~\citep{Sood2018}.  In general, a team choose to bat first expecting to set a high scoring target using superior field conditions in the first half of the match. The pitch condition is expected to deteriorate over time, which may eventually turn less favorable for batting in the second innings.  In contrast to this thought, the decision to bat first is also perturbed by a concern that a ‘safe’ score for confirming victory is unknown while batting in the first innings. Therefore, a team may choose to bowl first as they prefer to have a ‘known’ target score to chase expecting that field conditions or weather may rather turn unfavorable to bowlers in the second innings. Eventually, the coin toss decision inevitably tends to favor one team over another to some extent, which may not always give both teams a fair chance to win the match. Furthermore, statistics from the last four world cup cricket tournaments suggest that the teams winning the toss and deciding to bowl first are victorious in less than 50\% of similar cases~\citep{newsCon}. This phenomenon reveals certain advantages in batting first over bowling apparently due to a number of confounding factors.  In support to this observation, Dawson et al. have concluded in their study that winning the toss and batting first significantly increases the chance of winning the match compared to the decision of bowling first after winning the toss~\citep{Dawson2009}.

\begin{figure*}[t]
%\centering
\subfigure[] { \includegraphics[width=0.36\textwidth]{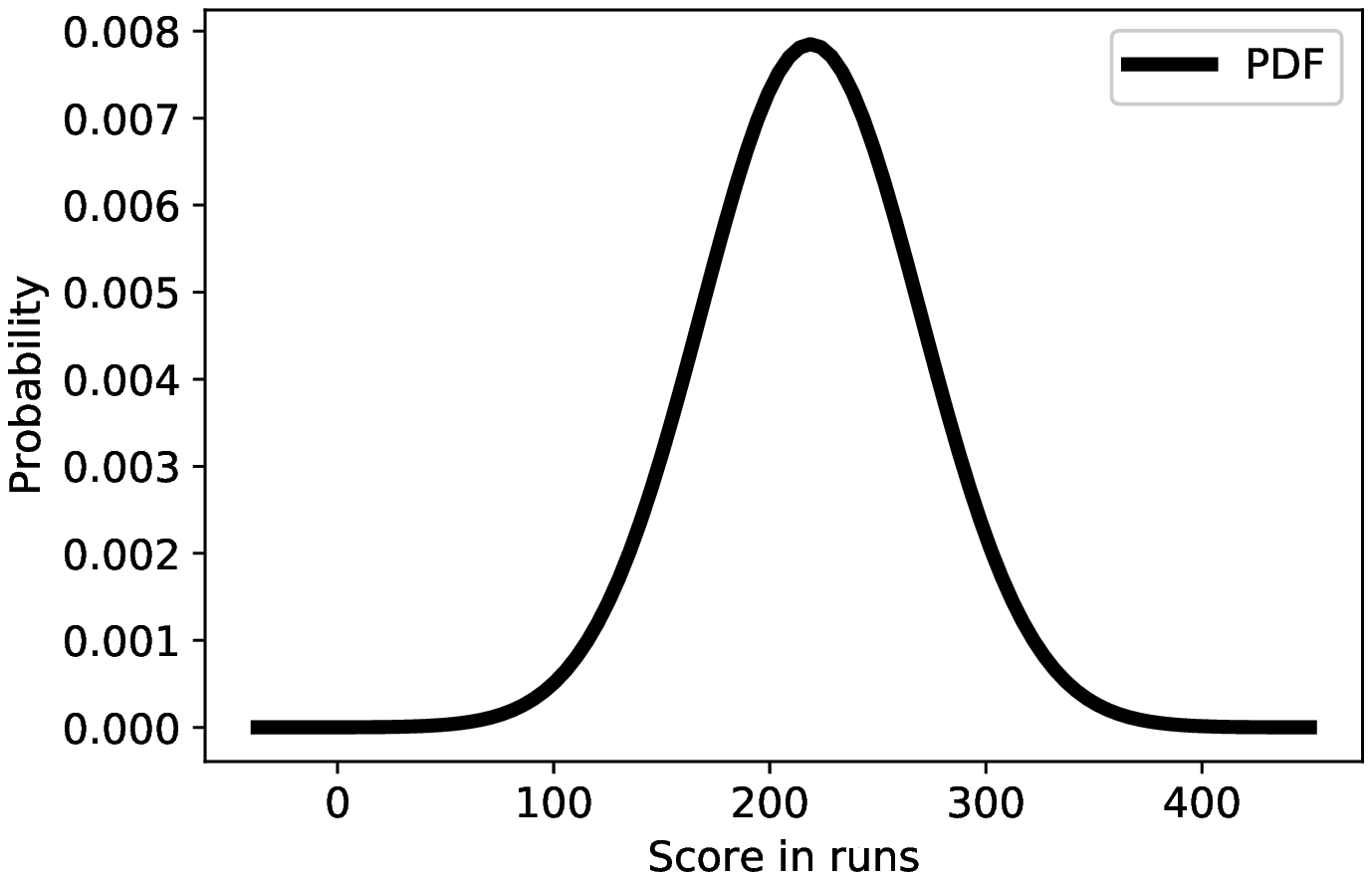}} %Figure/PDF.eps}}
\hspace{-22pt}
\subfigure[] { \includegraphics[width=0.36\textwidth]{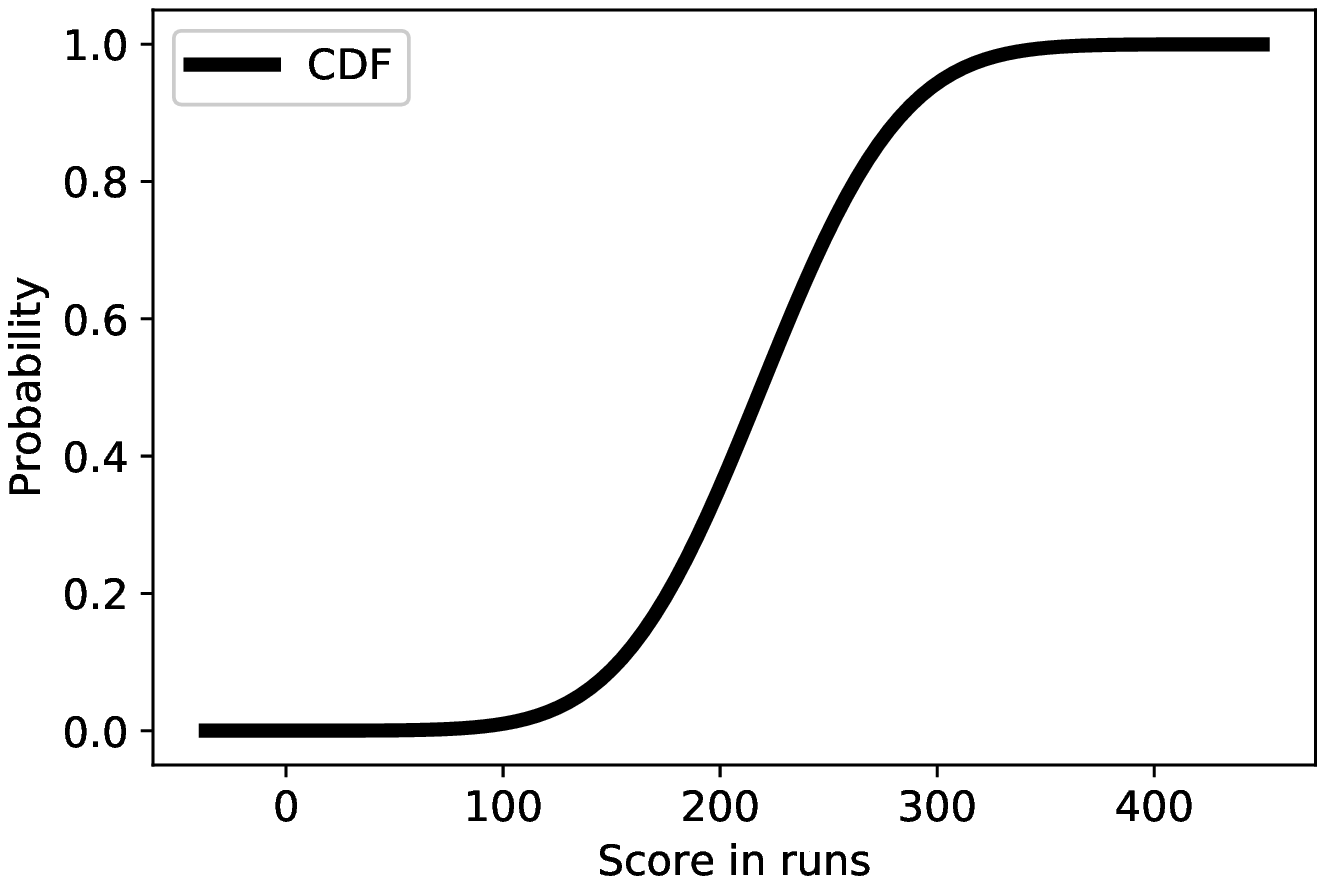}}% {Figure/CDF.eps}}
\hspace{-21pt}
\subfigure[] { \includegraphics[width=0.36\textwidth]{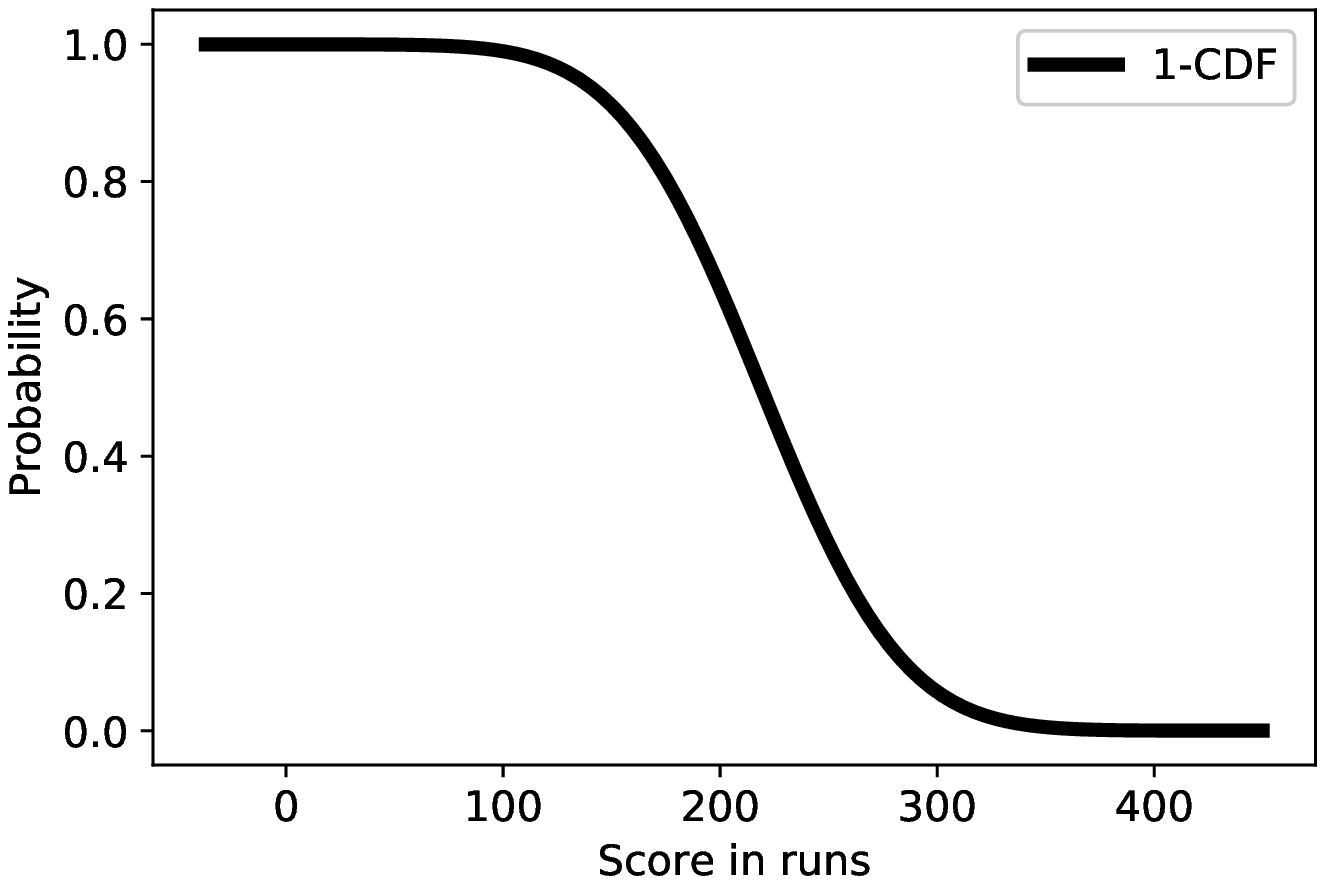}}%{Figure/1-CDF.eps}}
\caption{(a) Probability density function, P(x), (b) cumulative density function (CDF) of P(X), and (c) complement of CDF of scoring runs representing probability of scoring greater than X runs.} 
\label{fig01}
\end{figure*}

\subsection {Proposed research} 

In line with above observations, we identify two cases that may significantly bias the outcome of the game due to coin toss-based decision of batting or bowling order.  First, it is widely known that chasing a target score of over 300 runs in the second innings is often more challenging than scoring such high total in the first innings. This becomes more true even while chasing a decent target in the fourth innings of Test cricket. For example, there are only nine matches in over forty years of world cup cricket history where the bat-second team have been able to chase a target score of 300 and above runs. However, there are at least 19 matches in the 2019 world cup alone where the first innings scores are more than 300. Therefore, deciding to bat first and setting a big target appears to lower the probability of successfully chasing the target in the second innings even with two equally competitive teams. Second, first innings batting experience is free of pressure from chasing a target score, which is a psychological advantage. The team batting in the second innings incur this additional pressure and such pressure may negatively affect their natural batting performance. Even a very competitive team have been seen to collapse by scoring unusually low score while chasing an unusually high score~\citep{Bhattacharjee2016}, which is often termed as 'choking' or ‘strangling’ and has been studied by Lemmer~\citep{Lemmer2015}.  All these phenomena do not guarantee equal opportunity for teams to compete in a fair ground since the coin toss decision may ultimately influence the outcome of the game. These issues can negatively affect the excitement, competitiveness, and spirit of the game.

All these statistics raise a number of research questions that demand data-driven solutions. First, is there any run scoring advantage in the first innings score that might affect the outcome of the game? Second, what would the team score if they were to bat in the second innings given the fact that they have scored a big and ‘hard-to-chase’ total in the first innings? Third, what would be a reasonable and competitive target that accounts for the advantage of winning the toss and setting an unusually high first innings score? Since there is apparently no alternative to coin tossing, we have used probability theory on venue-specific ODI match results to investigate bias in run scoring distributions at different playing conditions and match outcomes. A statistical model has been proposed to incorporate such venue-specific bias for two purposes: 1) data mining to identify the magnitude of the bias and 2) recommending a revised target that will equalize the probability of winning a big scoring match regardless of the playing order (bat first versus bowl first). 

\section {Methodology}

This study proposes a probability-based model in investigating potential scoring bias following the decision of playing order taken by tossing a coin. First, the scoring probability distributions of four match cases are obtained and compared for each of ten ODI venues. The four match cases are: 1) bat-first-win, 2) bat-first-lose, 3) bat-second-win, 4) bat-second-lose. The run distributions are modeled using negative binomial (NB) distribution since it effectively represents run distributions in Scarf et al.~\citep{Scarf2011}. The NB distribution modeling a discrete random variable x is shown as below. 

%%%%%%%%%%%%%%%%%%
\begin {equation}
P(x,n,p) = \frac{\Gamma (x+n)}{\Gamma {(n)} x!} p^n (1-p)^x.
\end {equation}
%%%%%%%%%%%%%%%%%%%%
The parameters of NB distribution (n and p) are obtained using maximum likelihood estimates to yield the probability mass function (PMF). For comparison, probability density function (PDF) of contribution variable distributions are developed using mean and variance of the scores.  The PMF or PDF represent the scoring probability distribution, P(X). Figure 1(a) shows the PDF of a normal distribution. Intuitively the probability of scoring a total of exactly Xt runs P(X = Xt) (e.g. 237 runs) is low when the sample space is large typically ranging from 100 runs to 350 runs. Intuitively, scoring a total of 200 runs and more is much higher than scoring 300 runs and more. This leads to the definition of cumulative PDF that represents the probability of scoring up to Xt runs, P (X$\leq$Xt) by integrating or summing the PDF or PMF from zero to Xt, respectively as shown in Figure 1(b). We take complement of CDF in Eq. 1 to represent our intuition that scoring at least 200 runs P(X$>$200) is higher than that of scoring at least 300 runs, P(X$>$300), as shown in Figure 1(c).
%%%%%%%%%%%%%%%%%%
\begin{equation}
P (X>Xt) = 1 - P(X \leq Xt) = 1 – \sum_{x=0}^{x=xt} P(x,n,p)
\end{equation}
%%%%%%%%%%%%%%%%%
where, P(X,n,p) is the best fitted PMF on the data. We use the complement of CDF in subsequent comparisons among four match cases to investigate bias in scoring probability.

\subsection {Bayesian model}

In the presence of a scoring bias, we propose a Bayesian model to recommend a revised target score that will equalize the scoring and winning probability of both innings irrespective of the coin toss decision.  First, the variables and outcomes are identified for ODI cricket matches. The match outcome can be either win (W) or lose (L). First and second innings batting conditions are represented by BF and BS, respectively. We define the posterior probability of winning a match given that the team bat first and score at least Xf runs in the first innings as P (W $\mid$ S$>$Xf, BF). Using the Bayes' rule, this posterior winning probability can be calculated as below.

\begin{equation}
P (W \mid  S>Xf, BF) =  \frac{P (S>Xf, BF \mid  W)~P(W)} {P (S>Xf, BF)}  
\end{equation}

Similarly, the posterior probability of winning the match given that the winning team bat in the second innings and score Xs runs is as follows. 

\begin{equation}
P (W \mid S>Xs, BS) =  \frac{P (S>Xs, BS \mid W)~P(W)} {P (S>Xs, BS)}
\end{equation}
%%%%%%%%%
Given a first innings score of at least Xf runs, the minimum second innings target score Xs that will equalize the winning probability of any team in a particular venue is obtained by equalizing Eqs. 3 and 4 as below.

\begin{equation}
\frac {P(S>Xs, BS \mid  W)}{P(S>Xs, BS)} =  \frac {P(S>Xf,BF \mid  W)}{P (S>Xf,BF)} 
\end{equation}

Applying the chain rule of conditional probability,
%%%%%%
\begin{eqnarray}
\frac{P (S>Xs \mid  BS, W)~P(BS \mid W)} {P(S > Xs, BS)} =  \frac{ P (S>Xf \mid  BF,W)~P(BF \mid W)} {P (S>Xf, BF)} 
\end{eqnarray}
%%%%%%
Joint probability distributions in the denominator can be expressed in terms of conditional probability distributions. 
%%%%%%%
\begin{eqnarray}
%\centering
\frac{P (S > Xs \mid  BS, W)~P(BS \mid W) } {P (S>Xs \mid BS)~P(BS)} = \frac {P (S>Xf \mid  BF, W)~P(BF \mid  W) } {P (S>Xf \mid  BF)~P(BF)}
\end{eqnarray}
%%%%%%
\begin{eqnarray}
P (S>Xs \mid  BS, W) = \frac{P (S>Xs \mid  BS) } {P (S>Xf \mid  BF)}  \frac {P (BF \mid  W)} {P (BS \mid W) }~P(S>Xf \mid  BF, W).
\end{eqnarray}
%%%%%%
Here, the probability of batting first or second is equal, P(BF) = P(BS), considering the coin is fair. Given the first innings score Xs, we assume that the revised second innings score Xf will additionally equalize the scoring probability (in addition to the winning probability) of both innings such that 
P (S$>$Xf $\mid$ BF)= P (S$>$Xs $\mid$ BS). The revised equation is as follows. 
%%%%%%%
\begin{equation}
P (S>Xs \mid  BS, W) = C* P (S> Xf \mid BF, W)
\end{equation}
%%%%%%%%%%

Here, C = $\frac {P (Bf \mid  W) } {P (BS \mid  W) } $ is a constant ratio for a venue, which is the ratio of probabilities of batting first and second given that the team is victorious. In the case of higher likelihood of batting first of the winning team, the ratio will be greater than 1. Given the first innings score Xf, the second innings score Xs that satisfies the two conditions can be obtained by taking inverse of the CDF, P(S$\leq $Xs$\mid$BS,W) as below.
\begin{eqnarray}
Xs &=& \mbox{Inv}~(P (S\leq Xs \mid  BS,W)) \nonumber \\
 &= &\mbox{Inv}~(1- C* P (S>Xf \mid  BF, W)) 
\end{eqnarray}

\section {Results and Discussion}

We have implemented the entire model in Python programming language using the scipy, numpy, and pandas packages~\citep{McKinney2011pandas, Oliphant2007} and shared the source codes, data, and notebook in a github repository. 
Results of all ODI matches are obtained from the webpage of Cricket-stats~\footnote{http://cricket-stats.net/genp/grounds.shtml} for ten most popular international venues as summarized in Table~\ref{tab01}. Table~\ref{tab01} shows that the bat-first-lose teams have higher average score than that of bat-second-lose teams.

\begin{figure*}[t]
\centering
\includegraphics[width=0.7\textwidth]{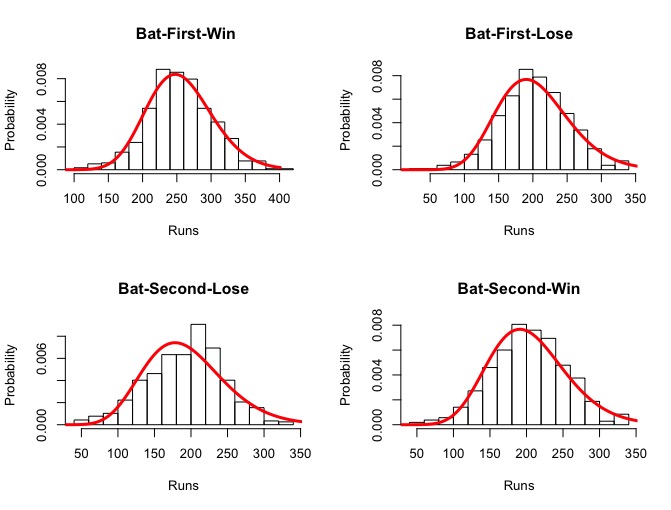}%{Figure/Fig_2_Rplot.jpeg}
\caption{Fitting runs of four experimental cases using the negative binomial distribution.} 
\label{fig02}
\end{figure*}

\begin{table*}[t]
\centering
\caption{Summary of one-day international cricket match results used in this study. Average runs are rounded off to the next integer value. Since Bat-second-win teams score as much as Bat-first-lose teams, their run scoring distributions appear similar.}
{\begin{tabular}{lccccccc} \toprule
Venue &Total match &\multicolumn{2}{l}{Bat-first-win}& Bat-second-lose&\multicolumn{2}{l}{Bat-second-win}& Bat-first-lose\\  \cmidrule{3-8}
& & \% & Avg. score & Avg. score & \% & Avg. score & Avg. score \\\bottomrule
Auckland	&71&	42.3&	240&	185	&57.7&	200&	203\\
Bangalore	 & 22	&50.0	&294	&248	&50.0	&236	&234\\
Harare	&149 &	49.7	&255&	183&	50.3&	205&	204\\
Lahore	&58	&56.9	&266	&205	&43.1	&233	&231\\
Lords	&62&	48.4	&268&	215	&51.6	&218&	217\\
Melbourne	&145	&49.7	&245	&191	&50.3	&202	&201\\
Mirpur	&107&	46.7	&261&	194	&53.3&	203&	204\\
Premadasa	&118&	58.5	&266	&196	&41.5	&204	&203\\
Sharjah	&236	&53.8&	252&	189&	46.2&	195&	192\\
Sydney	&149	&59.1&	248&	189	&40.9&	195	&198\\ \bottomrule
Overall	&1117	&51.5&	260	&200&	48.5&	209&	209\\ \bottomrule
\end{tabular}}
\label{tab01}
\end{table*}

\subsection {Analysis of probability distributions}

The probability distribution of run scored at each venue is studied by fitting the negative binomial distribution. The NB distribution is one of the popular choices for modeling count data like runs in Cricket. The NB distribution is also fitted to yield the overall distribution of runs regardless of the venue. The effect of batting and bowling order on scoring distribution is analyzed by categorizing the distributions into four cases: 1) bat-first-win, 2) bat-second-lose, 3) bat-first-lose, and 4) bat-second-win. Figure~\ref{fig02} shows scoring distributions of four experimental cases after fitting with NB distribution using all venue data. Following this result, we assume that venue-specific distributions also follow the NB distribution.

\begin{table*}
\centering
\caption{Revised second innings target score against a first-innings score (actual score). The revised scores are shown for binomial distribution. The sum of difference between actual and revised scores are shown for the three distributions: negative binomial, normal, and logistic. The overall model included data from all 1117 ODI matches and is not mere aggregation of ten venue results.}
{\begin{tabular}{lcccccccc} \toprule
Actual Target &300&	315	&330	&340&	350&\multicolumn{3}{c}{Total difference in runs}\\ \cmidrule{2-6}
Venue &\multicolumn{5}{c}{Negative binomial distribution}& Negative binomial& Normal  & Logistic \\ \midrule
Auckland	&283&	301&	320&	332&	345&	54	&63&	6\\
Bangalore&	241&	251	&261	&268	&275&	335	&329	&327\\
Harare&	276	&289&	312	&327	&342	&95&	155	&152\\
Lahore&	251&	266	&280	&289&	298&	250&	233&	283\\
Lords&	263&	284	&306	&321	&335	&120	&145	&134\\
Melbourne	&267&	286	&304	&317&	329	&131	&164	&163\\
Mirpur	&255	&273&	291&	303&	316&	197&	211&	184\\
Premadasa&	225	&243	&260	&271	&282	&355	&363	&420\\
Sharjah	&242	&259&	277&	289&	301&	267&	282&	308\\
Sydney	&230	&247&	263	&274	&285	&336	&343	&414\\
Overall model	&249&	266&	284&	295&	307&	234&	245&	261\\ \midrule
\end{tabular}}
\label{tab02}
\end{table*}

\begin{figure*}
\centering
{ \includegraphics[width=0.4\textwidth]{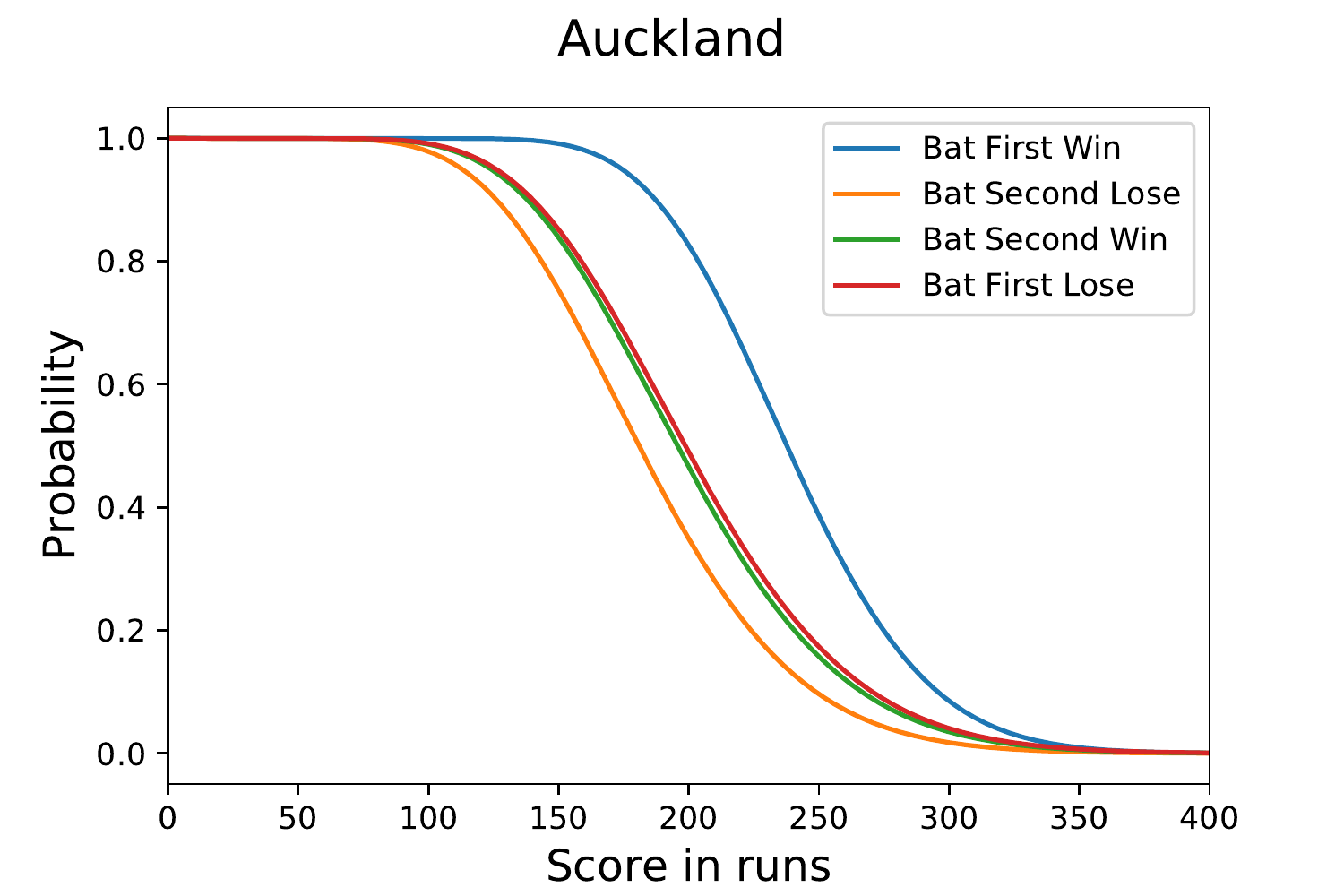}}
{ \includegraphics[width=0.4\textwidth]{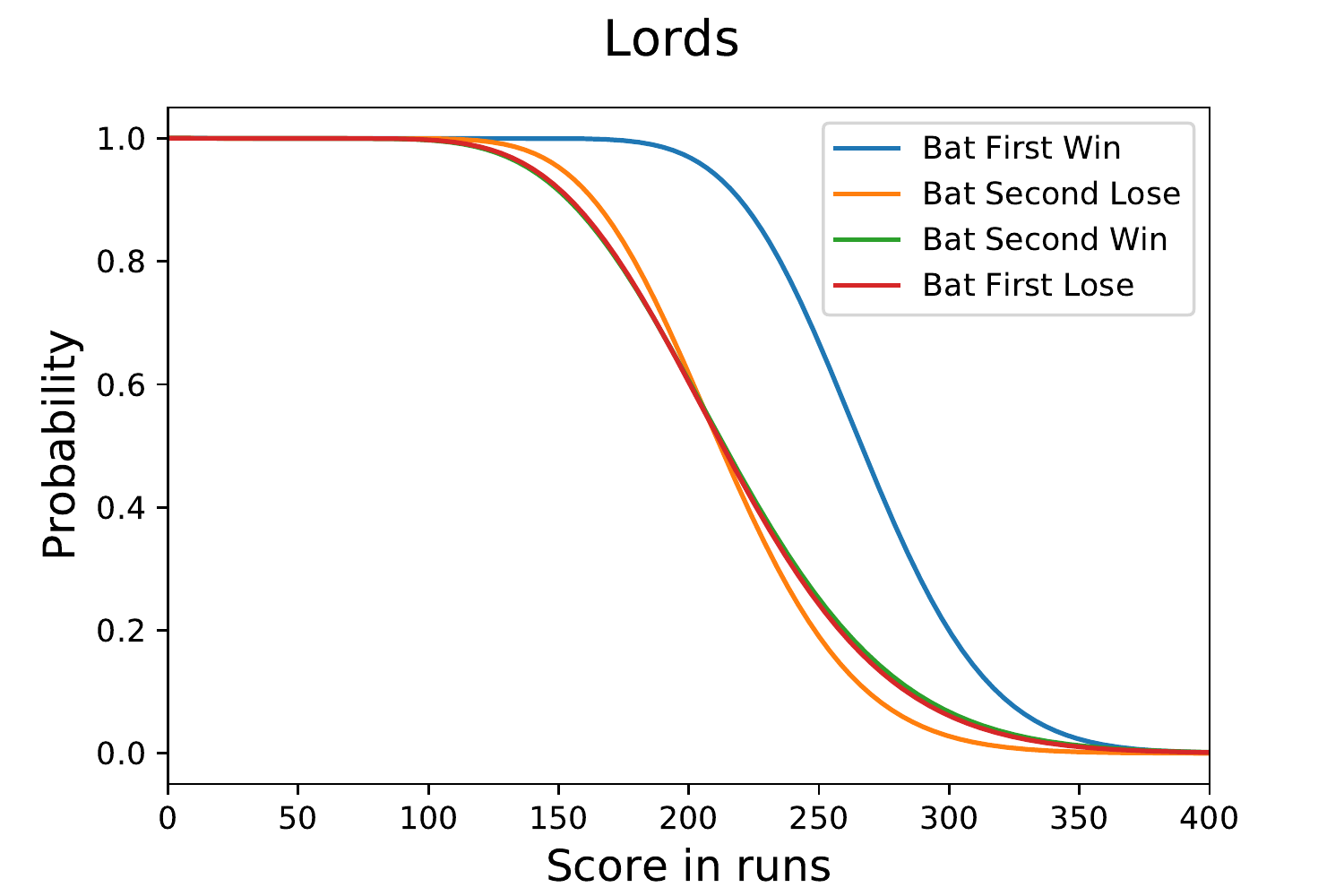}}
{ \includegraphics[width=0.4\textwidth]{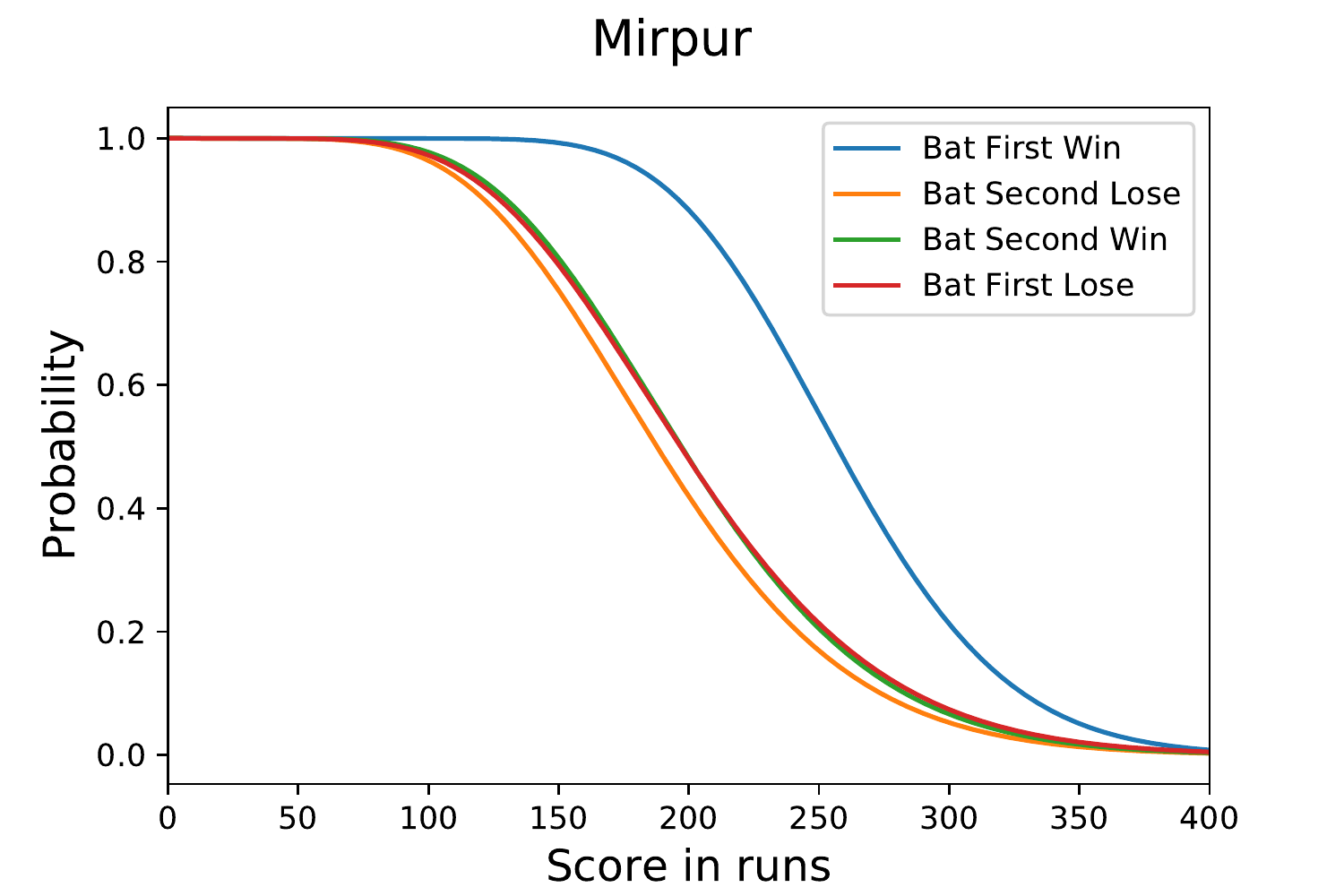}}
{ \includegraphics[width=0.4\textwidth]{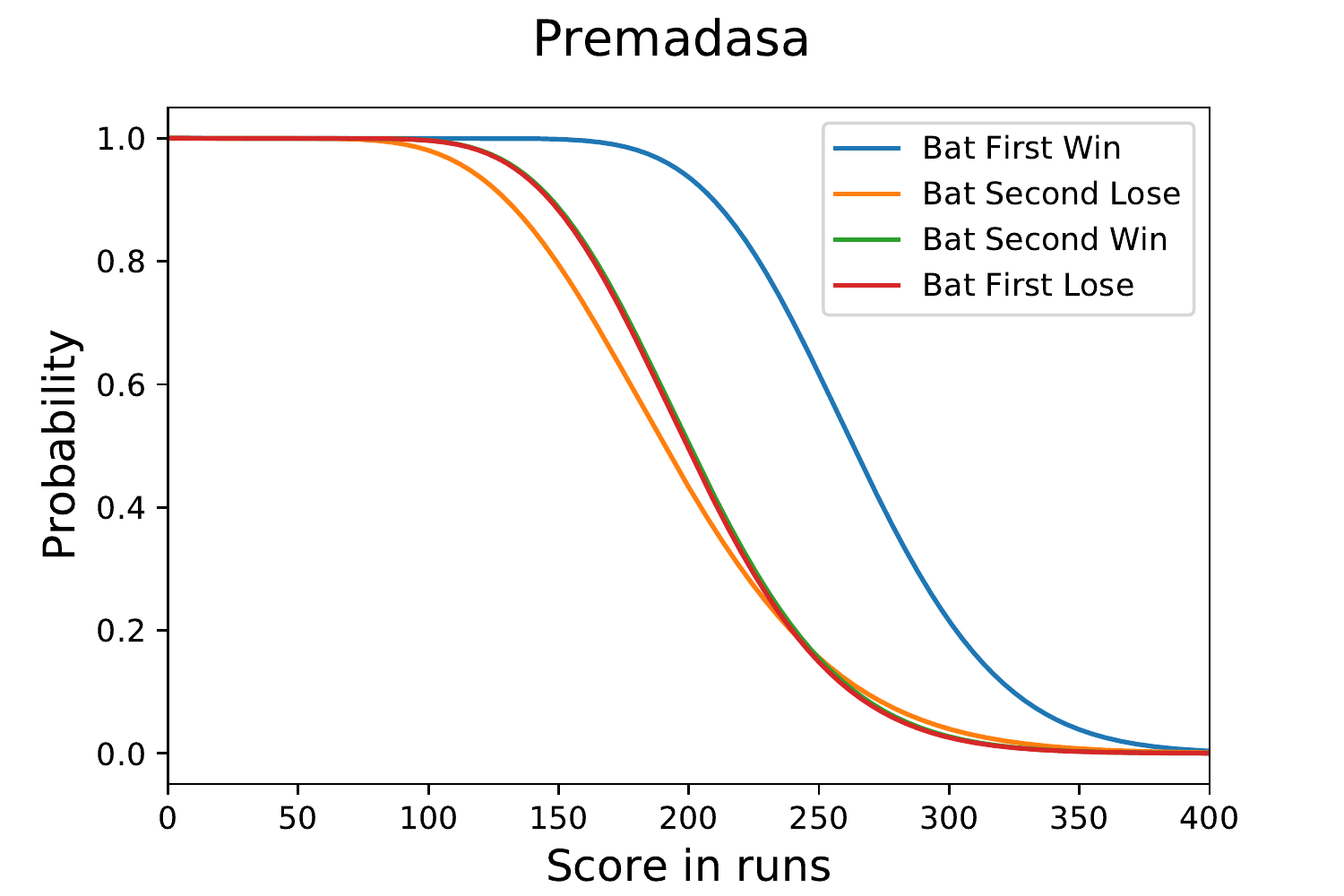}}
{ \includegraphics[width=0.4\textwidth]{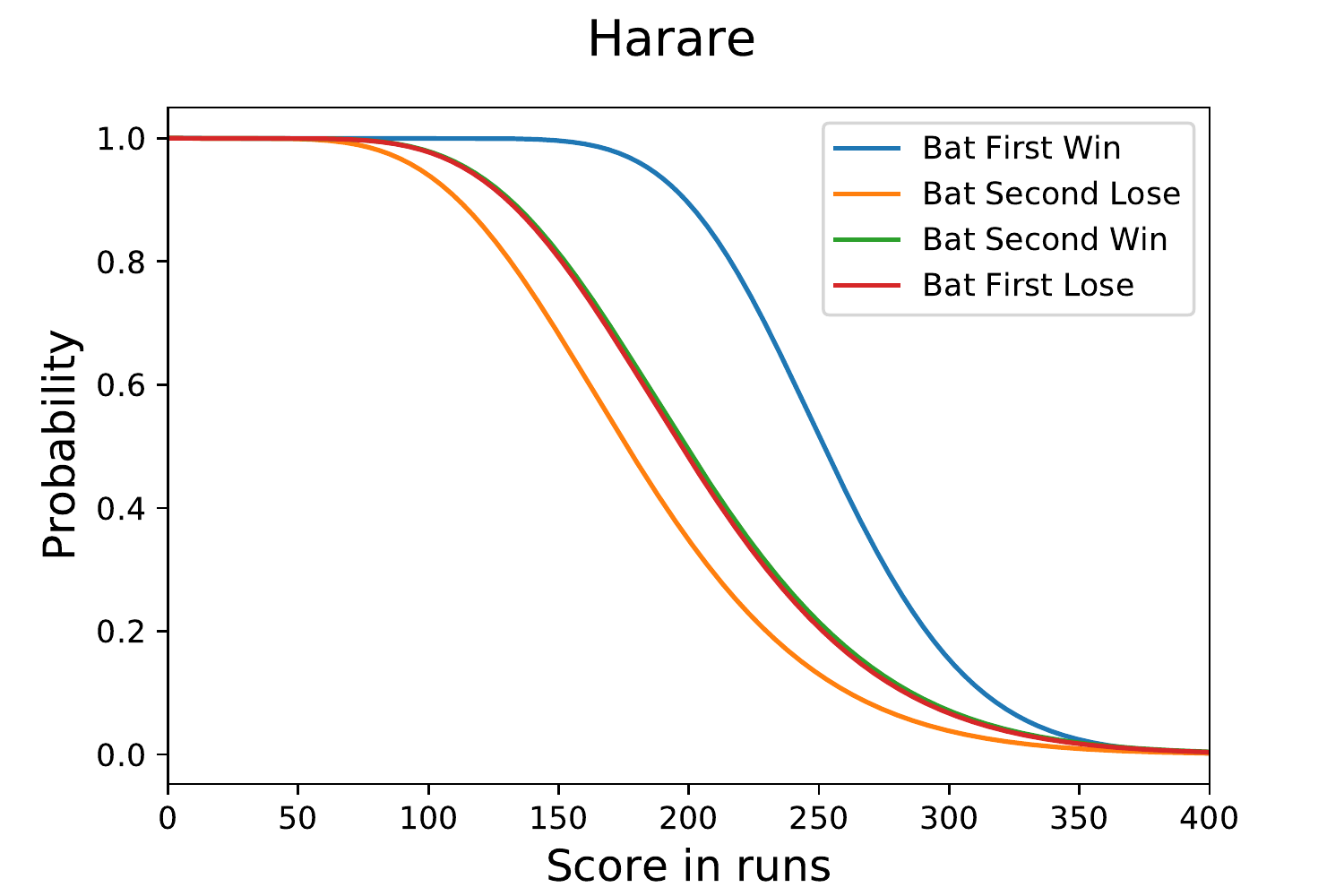}} 
{ \includegraphics[width=0.4\textwidth]{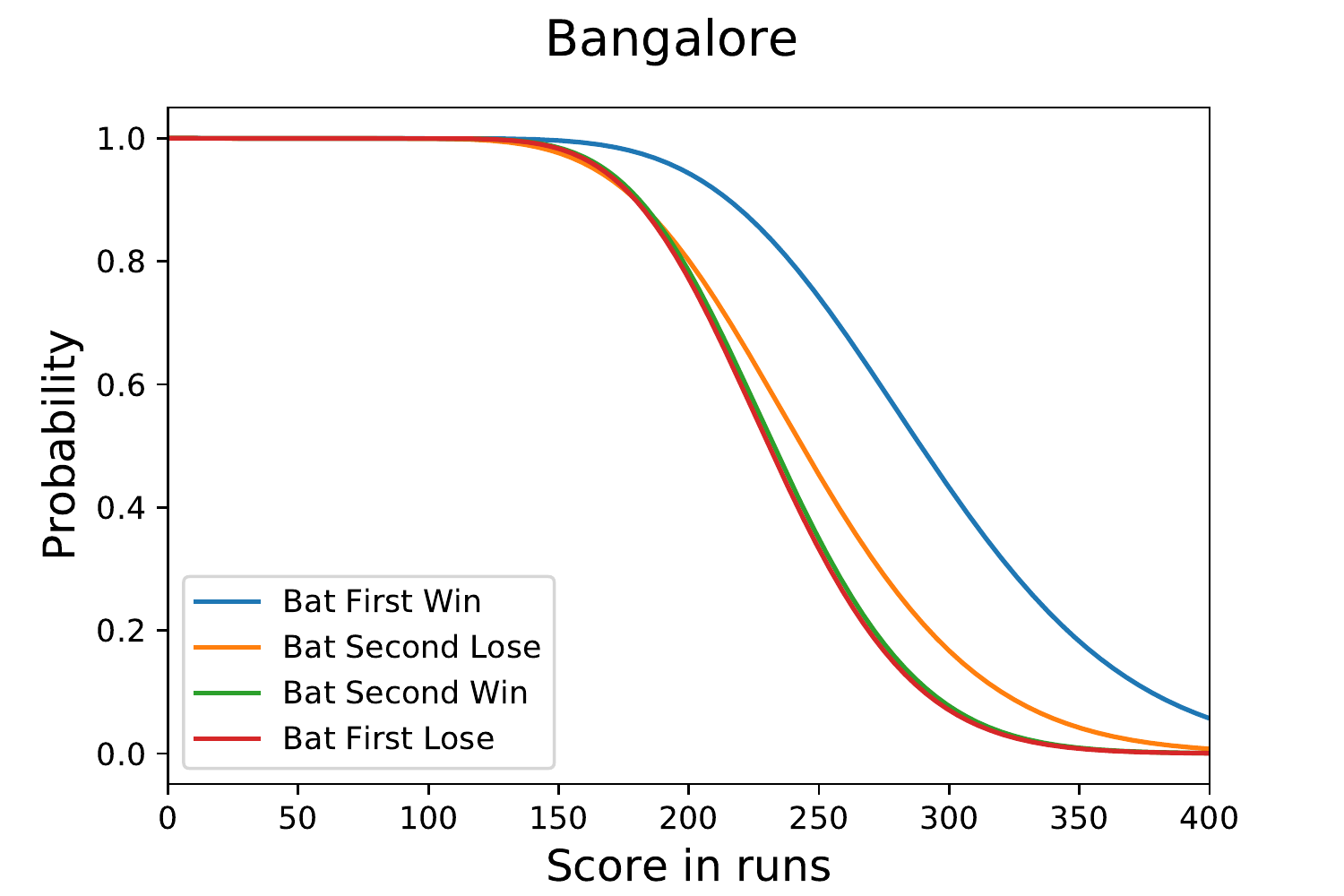}}
{ \includegraphics[width=0.4\textwidth]{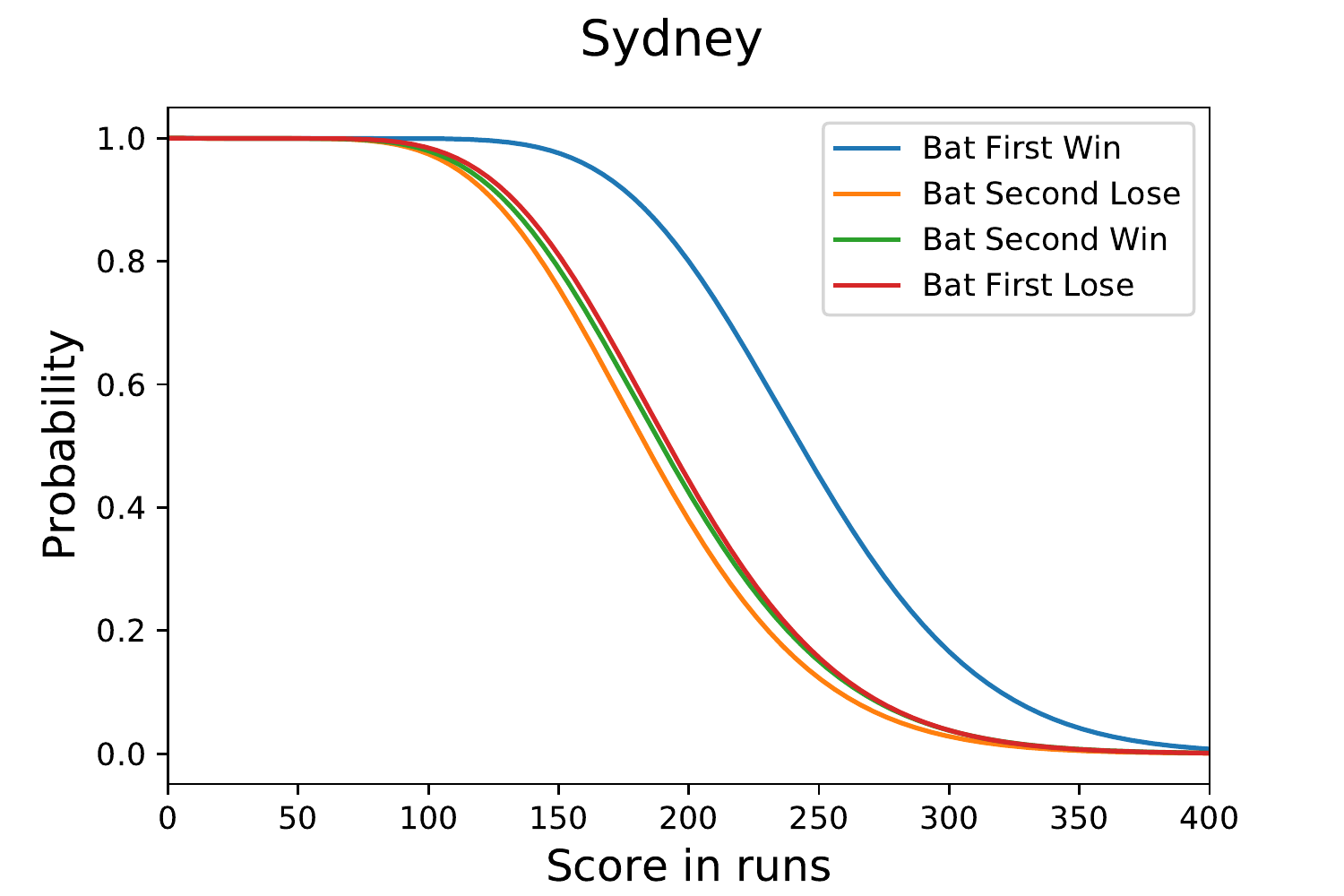}}
{ \includegraphics[width=0.4\textwidth]{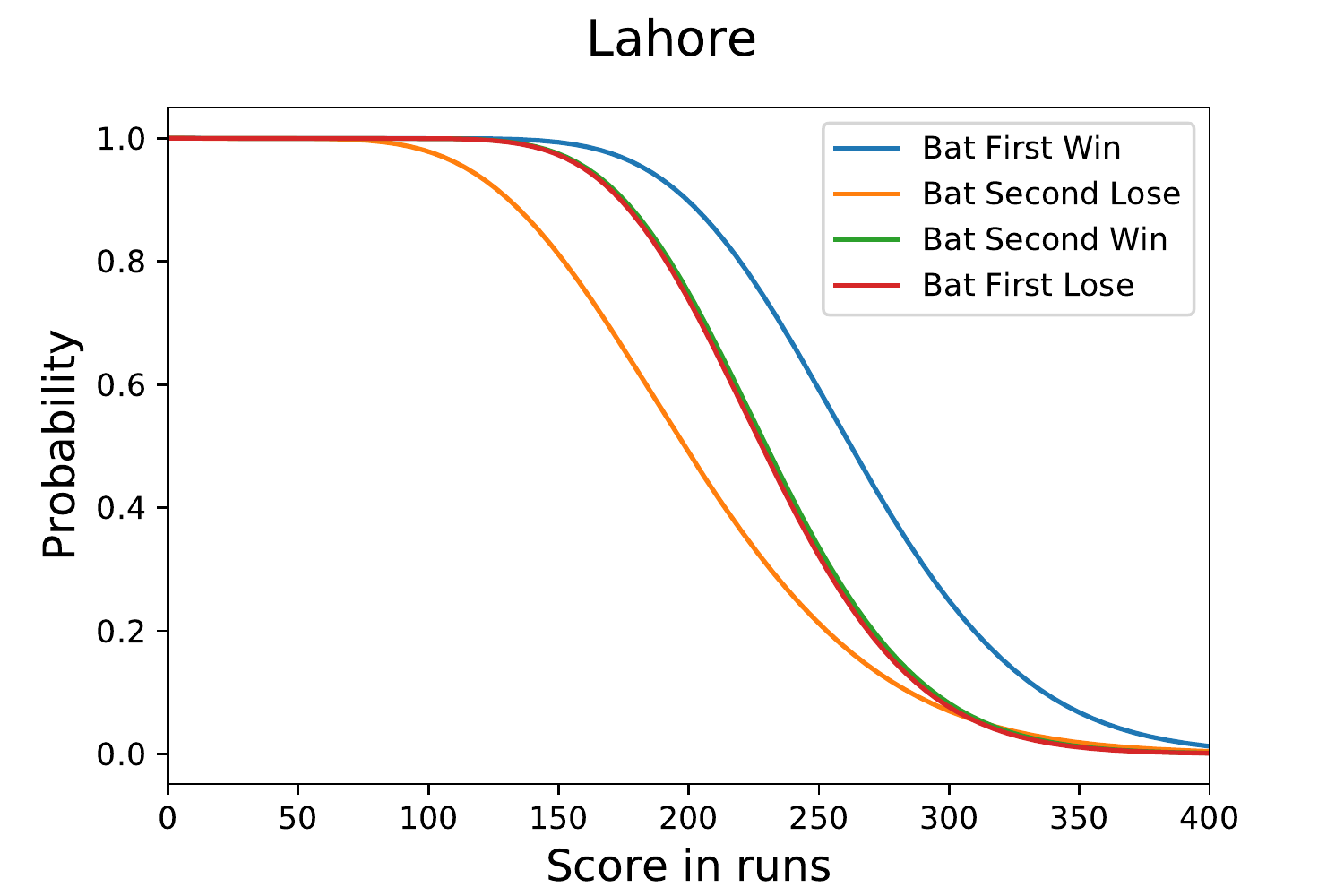}}
 { \includegraphics[width=0.4\textwidth]{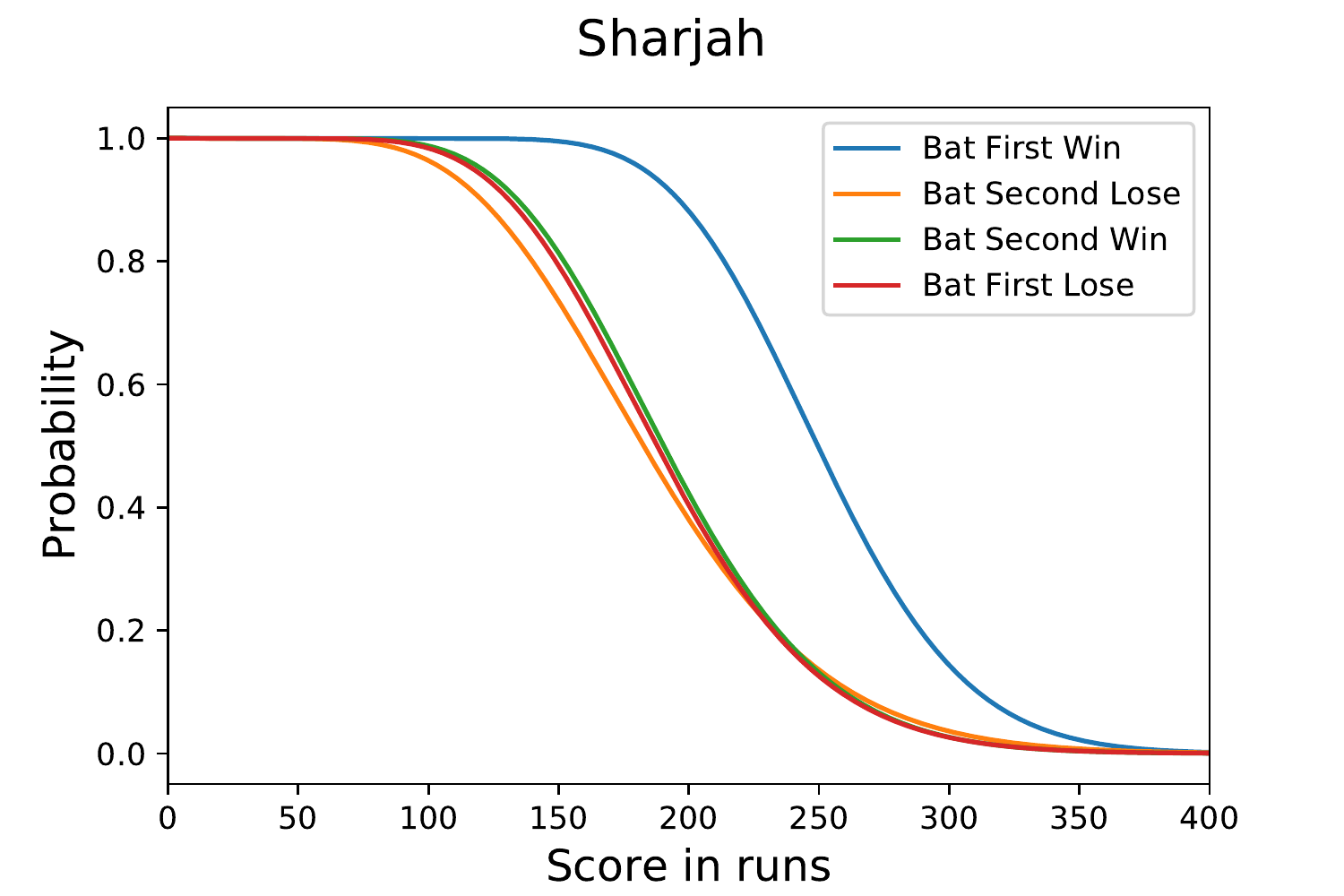}}
  { \includegraphics[width=0.4\textwidth]{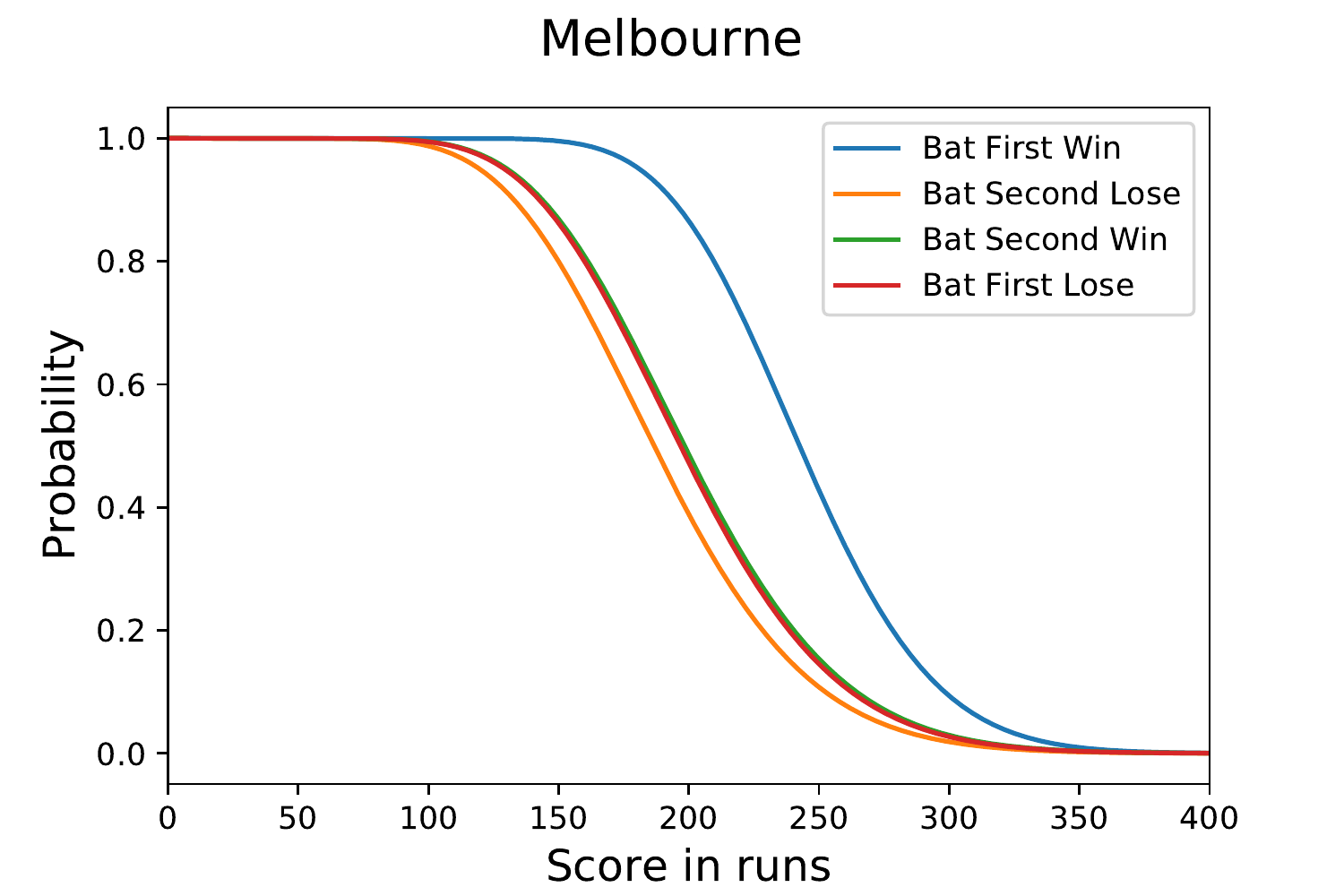}}
\caption{Complement of cumulative negative binomial distribution of four cases: two playing orders (Bat first, Bat Second) and two outcomes (win, lose) at ten ODI venues. All venues show large discrepancy between scoring probabilities of 'bat-first-win' teams and their opponent 'bat-second-lose' teams.} 
\label{fig03}
\end{figure*}

Figure~\ref{fig03} shows that the run distributions of bat-first-lose and bat-second-win are similar and overlapping for all venues. This is intuitive since the bat-second-win team will always score one or several runs more than the bat-first-lose team. These two cases have yielded same average score of 209 runs over 1117 ODI matches (Table~\ref{tab01}). However, the scoring distribution of bat-second-lose teams falls behind that of bat-first-lose teams for almost all venues. That is, batting first yields higher scoring probability than that of batting in the second innings even when both cases the team lose the match. This discrepancy is indicative of a scoring bias or advantage possibly due to the batting order. This advantage in run scoring is more evident at venues like Premadasa, Harare, Lahore and lowest at Sharjah, Sydney, and Lords. The only exception is the venue at Bangalore where the bat-second-lose scoring probability is better than that of the bat-first-lose case. This observation may be attributed to the lowest sample size for Bangalore (only 22 ODI matches grouped into four categories) compared to other venues (Table~\ref{tab01}).

However, this is not the case when the bat-first team win the match against the bat-second-lose team. The data reveal that the 'bat-second-lose' teams have the worst scoring performance and their opponent 'bat-first-win' teams have the best scoring distribution among all four cases. The scoring distribution of bat-first-win teams is far ahead of those of other three cases for all venues (Figure~\ref{fig03}). This distribution also infers that the same high scoring team in the first innings is likely to score lower in the second innings. This suggests an extra disadvantage of batting in the second innings to chase a large target set by the bat-first team. It is noteworthy that these discrepancies are venue-specific under the assumption that stronger and weaker teams have similar likelihood of winning the toss or batting in the first innings. Such discrepancies can be even more striking when the stronger side gets the chance to bat first and sets a large and unattainable target to chase. Therefore, it is worth determining the total runs that the bat-first-win team would have scored if they were sent to bat in the second innings. This revised score may inform us about the extent of bias and recommend a fair target to be chased in the second innings to alleviate the effect of coin toss on match outcome. 
%%%%%%%%
\subsection {Probabilistic model for revising target}
The previous section reveals higher scoring probability of bat-first-win teams compared to that of any other three cases.  This high scoring probability can yield a target score that becomes very challenging to chase while batting in the second innings. For example, the world cup record of chasing the highest score is 329 runs whereas first innings score has been as high as 397 runs. There are at least eight matches in the 2019 cricket world cup alone with over 330 runs scored in the first innings, which would require the opponent to break the world cup record to win those matches. Statistics suggest that there are only a handful cases in the world cup cricket when a score of 300 runs and above is successfully chased in the second innings. Therefore, after observing a very high score in the first innings, the game is assumed over even before playing the second innings of the match. Commentary is made highlighting that there is no record of chasing such a high score in this particular venue. These observations go against the 'glorious uncertainties' of the game of cricket.   

One naive way to tackle this bias is to equalize the opportunity for both teams in a big scoring match by revising an unusually big target score set in the first innings considering a venue-specific or overall probabilistic model. In this study, we assume that any score over 300 runs is typically a tough score to chase in the second innings.  A revised target score for the second innings may be obtained using Eq. 10 that equalizes both winning and scoring probability of the competing teams. Table~\ref{tab02} shows revised scores for a set of first innings scores over 300 runs using the proposed probabilistic model.  The revised scores obtained from the normal and logistic probability distribution are not too different from those obtained from the negative binomial distribution.  The total run difference between the actual target (first innings score) and the model revised target is also a measure of bias for each venue. Results suggest that the venue in Auckland has the lowest difference and the one in Premadasa has the highest differences for all three distribution models. The negative binomial distribution has yielded the lowest differences in scoring among the three distributions because it is known to best fit run scoring data. The last row of Table~\ref{tab02} shows the overall model results obtained after fitting all 1117 match results from all ten venues. In all cases, the revised target score for the second innings team is lower than that of the first innings score.  
%%%%%%%
\subsection {Limitations}
The goal of this study is to investigate run scoring advantages due to batting in the first innings of cricket. The proposed Bayesian model makes two assumptions to determine the revised score to alleviate the bias present in the first innings score. The model equalizes both winning and scoring probabilities of a first innings score with those of a revised second innings score. However, the assumption of equal winning probability, when the scoring probability is considered unequal (P(S$>$Xf $\mid$ BF) $\neq$ P(S$>$Xs $\mid$ BS)), complicates the solution and does not yield a robust solution of the revised score. A more conservative estimate of the revised target (closer to the first innings score) may be obtained by relaxing one of the two assumptions.  Our model needs further search for robust solutions in providing more unbiased target scores by relaxing one of the two model assumptions. A number of venues have comparatively low sample size, which is further reduced due to grouping of the samples into four match cases. Therefore, venue-specific models may not be reliable when the sample size is low. Conversely, there may be high variance in the model that includes all venue data.  The proposed model is not directly applicable for new venues unless past data are combined from all other existing venues in the model development. There are no gold standard and benchmark data to evaluate the performance of the proposed model because of its empirical nature.

\section {Conclusions}
This study has investigated run scoring distributions of different playing orders and outcomes in the one-day international cricket.  Our data clearly suggest an extra advantage of batting first in high scoring matches regardless of venues and strength of the playing teams. The high scoring distribution of all bat-first-win cases also infers that the same bat-first-win team is highly likely to score less if they were sent to bat in the second innings to chase the same score. Our proposed Bayesian model has captured venue-specific run scoring distributions to show the magnitude of advantage in the first innings batting for the winning team and to estimate revised target scores for the second innings. The revised target scores ensure equal winning and scoring probability for a particular venue.  Despite limitations in numerical computations, we believe that this is one of the first studies to investigate such bias in the game of cricket with a recommended solution. The proposed model can be used to investigate ordering bias in other research and operations to subsequently recommend a revision to alleviate effects of any factor biasing the outcome.

\bibliography{cricket}

\end{document}